\documentclass[a4paper,11pt]{article}

\usepackage{amsmath}
\usepackage{amssymb}
\usepackage[T1]{fontenc}
\usepackage[a4paper,left=1.0in, right=1.0in,top=1.1in, bottom=1.2in]{geometry}
\usepackage{graphicx}

\usepackage{epsfig,bm}

\usepackage{jheppub} 

\usepackage[T1]{fontenc} 

\newcommand{\beq}{\begin{equation}}
\newcommand{\eeq}{\end{equation}}
\newcommand{\bea}{\begin{eqnarray}}
\newcommand{\eea}{\end{eqnarray}}

\title{\boldmath Single bottom quark production in $k_\perp$-factorisation}

\preprint{IFIC/15-43}

\author[a]{Grigorios Chachamis}
\author[b]{\hspace{-0.35em}, Michal De\'ak}
\author[c]{\hspace{-0.35em}, Martin Hentschinski}
\author[b]{\hspace{-0.35em}, Germ\'an Rodrigo}
\author[a]{and Agust\'in Sabio Vera}

\affiliation[a]{Instituto de F\'isica Te\'orica UAM/CSIC \& Universidad Aut\'onoma de Madrid, C/ Nicol\'as Cabrera 15, E-28049 Madrid, Spain}
\affiliation[b]{Institut de F\'isica Corpuscular, Universitat de Val\`encia -- Consejo Superior de Investigaciones Cient\'ificas, Parc Cient\'ific, 46980 Paterna, Val\`encia, Spain}
\affiliation[c]{Instituto de Ciencias Nucleares, Universidad Nacional Aut\'onoma de M\'exico, Apartado Postal 70-543, M\'exico Distrito Federal 04510, M\'exico}

\emailAdd{grigorios.chachamis@csic.es}
\emailAdd{michal.deak@ific.uv.es}
\emailAdd{martin.hentschinski@gmail.com}
\emailAdd{german.rodrigo@csic.es}
\emailAdd{a.sabio.vera@gmail.com}

\abstract{ We present a study within the $k_T$-factorisation scheme on single bottom quark production at the LHC. In particular, we calculate the rapidity and transverse momentum differential distributions for single bottom quark/anti-quark production. In our setup, the unintegrated gluon density is obtained from the NL$x$ BFKL Green function whereas we included mass effects to the L$x$ heavy quark jet vertex. We compare our results to the corresponding distributions predicted by the usual collinear factorisation scheme. The latter were produced with {\sc Pythia} 8.1.}

\begin{document}

\maketitle


\flushbottom

\section{Introduction}

Major theoretical developments in the last three decades in small-$x$
physics made possible phenomenological analyses of high energy
scattering processes within the $k_T$-factorization scheme
~\cite{Catani:1990xk,Catani:1990eg,Catani:1994sq} at $e p$ (HERA) and
hadron colliders (Tevatron, LHC).  The Balitsky-Fadin-Kuraev-Lipatov
(BFKL) framework for the resummation of high center-of-mass energy
logarithms at leading (LL)~\cite{BFKLLO} and next-to-leading
(NLL)~\cite{BFKLNLO} logarithmic accuracy is in the core of the
majority of these analyses.

It is very natural to wonder whether the knowledge acquired from the
study of Deep Inelastic Scattering (DIS) processes at HERA within the
BFKL formalism, mainly from the description of $F_2$ and $F_L$ data,
could be of direct use for the description of processes at the LHC. In
principle, factorization and universality dictate the existence of a
transition approach from $e p$ to hadron-hadron collisions
~\cite{Mueller:1986ey,Vera:2007kn,Kwiecinski:2001nh}, despite the
different kinematic phase space limits.  A simple way for that to be
realized and act as a proof of concept is to use an unintegrated gluon
density from HERA fits into a phenomenological study of an LHC
process.  Recently, there were successful attempts for the detailed
description of the $Q^2$ and $x$ dependence of the structure functions
$F_2$ and $F_L$ by making use of a collinearly-improved BFKL equation
at next-to-leading logarithmic (NL$x$)
accuracy~\cite{Hentschinski:2012kr,Hentschinski:2013id}.\footnote{See also the works in Refs.~\cite{Kowalski:2011zza,
    Kowalski:2010ue}.}

Within high energy factorization, the description of any hard process
requires three ingredients: the universal BFKL gluon Green's function
which resums high energy logarithms and two process dependent impact
factors which describe the coupling of scattering particles to the
gluon Green's function. In the present case, only one impact factor
(the `heavy quark impact factor') is characterized by a hard scale
{\it i.e.} the heavy quark mass and large transverse momentum which
enables us to calculated it using perturbative QCD and collinear
factorization. The second impact factor (the `proton impact factor'),
which describes the coupling of the gluon Green's function to the
proton, is intrinsically non-perturbative and needs to be
modeled. Combination of the gluon Green's function and the proton
impact factor yields then the above mentioned unintegrated gluon
density.  The impact factors for gluons and massless quarks have been
calculated in
Ref.~\cite{Ciafaloni:1998hu,Hentschinski:2011tz,Chachamis:2012cc}, at
NL$x$. The NL$x$ impact factor for a massive quark in the initial
state has been calculated in
Ref.~\cite{Ciafaloni:2000sq,Chachamis:2013bwa}.

In the last years, studies of BFKL evolution 
were mainly focused on processes with two hard scales of similar sizes
in the final state to suppress any collinear-like evolution, with
Mueller-Navelet jets~\cite{Mueller:1986ey} the best known
example. Most of the studies were carried out at NL$x$
accuracy~\cite{Caporale:2014gpa, Ducloue:2013bva, Ducloue:2013wmi,
  Caporale:2012ih, Colferai:2010wu, Marquet:2007xx,Vera:2007kn,
  Angioni:2011wj,Caporale:2013uva}.

On the other hand there has been also considerable interest in the
study of processes with one hard scale, which involve unintegrated or
Transverse Momentum Dependent (TMD) parton density functions (PDFs). Examples of such processes
at the LHC include forward jet
~\cite{Chachamis:2012mw,Hentschinski:2014lma, Deak:2009xt,
  Deak:2011ga} and forward Z production~\cite{Hautmann:2012sh, Dooling:2014kia}. During
recent years the study of TMD PDFs has become a very active area of
research, which find applications in various multi-scale processes in
hadronic collisions, see Ref. \cite{Angeles-Martinez:2015sea} for a
recent review. Extraction of TMD PDFs has in some cases been developed
to very sophisticated levels, including a detailed discussion of
experimental uncertainties, see {\it e.g.}~\cite{Alekhin:2014irh}.

In this paper, we study single bottom (or anti-bottom) quark
production at the LHC. Bottom quark production (more accurately,
bottom pair production) has received lots of attention in the
literature~\cite{Gauld:2015yia,Nason:1987xz,Berger:1997gz,Bonciani:2009nb,Bonciani:2010mn,Fujii:2013gxa,Luszczak:2013cba,Fujii:2013yja,Zhu:2013yxa,Bonciani:2013ywa,Kniehl:2008fd,Berger:1993yp}
both in the collinear and the $k_T$-factorization approach. Bottom
quarks can generally be produced via gluon splitting, $g \rightarrow b
\overline{b}$ in proton-proton collisions.  Since our main purpose
here is to test the unintegrated gluon density from the HERA fit
\cite{Hentschinski:2012kr,Hentschinski:2013id} and to compare it to
theoretical predictions from collinear factorization at small $x$, we
concentrate in the following on bottom quark production in the forward
region of one of the protons. In this way the heavy quark -- as an
incoming parton -- will be fixed at relatively large $x$, while the
second parton -- a gluon -- is forced into the small-$x$ region.
Measurement of such a process will be possible within the LHCb
experiment~\cite{Alves:2008zz} and currently discussed forward updates
of the ATLAS and CMS experiments.

While the unintegrated gluon density extracted from
\cite{Hentschinski:2012kr,Hentschinski:2013id} does not provide a
detailed discussion of experimental uncertainties (unlike {\it e.g.}
\cite{Alekhin:2014irh}), it is the only currently available
unintegrated gluon density which is subject to BFKL evolution at NLL
accuracy, including a resummation of large logarithms at the level of
the next-to-leading order BFKL kernel. In this sense the
current studies present an advance over previous attempts based on LL
accuracy.

The article is structured as follows: in Section~\ref{sec:hef} we
introduce the high energy factorization framework we will use and in
Section~\ref{sec:diff-cs} we derive the master formula for the
differential single bottom quark cross-section.  In
Section~\ref{sec:numres} we present the numerical results and we
conclude in Section~\ref{sec:conclusions}.

\section{Forward single bottom quark production in high energy factorization}\label{sec:hef}  

\begin{figure}[tbh]
\vspace{0cm}
  \begin{picture}(0,0)
    \put(40, -220){
      \includegraphics{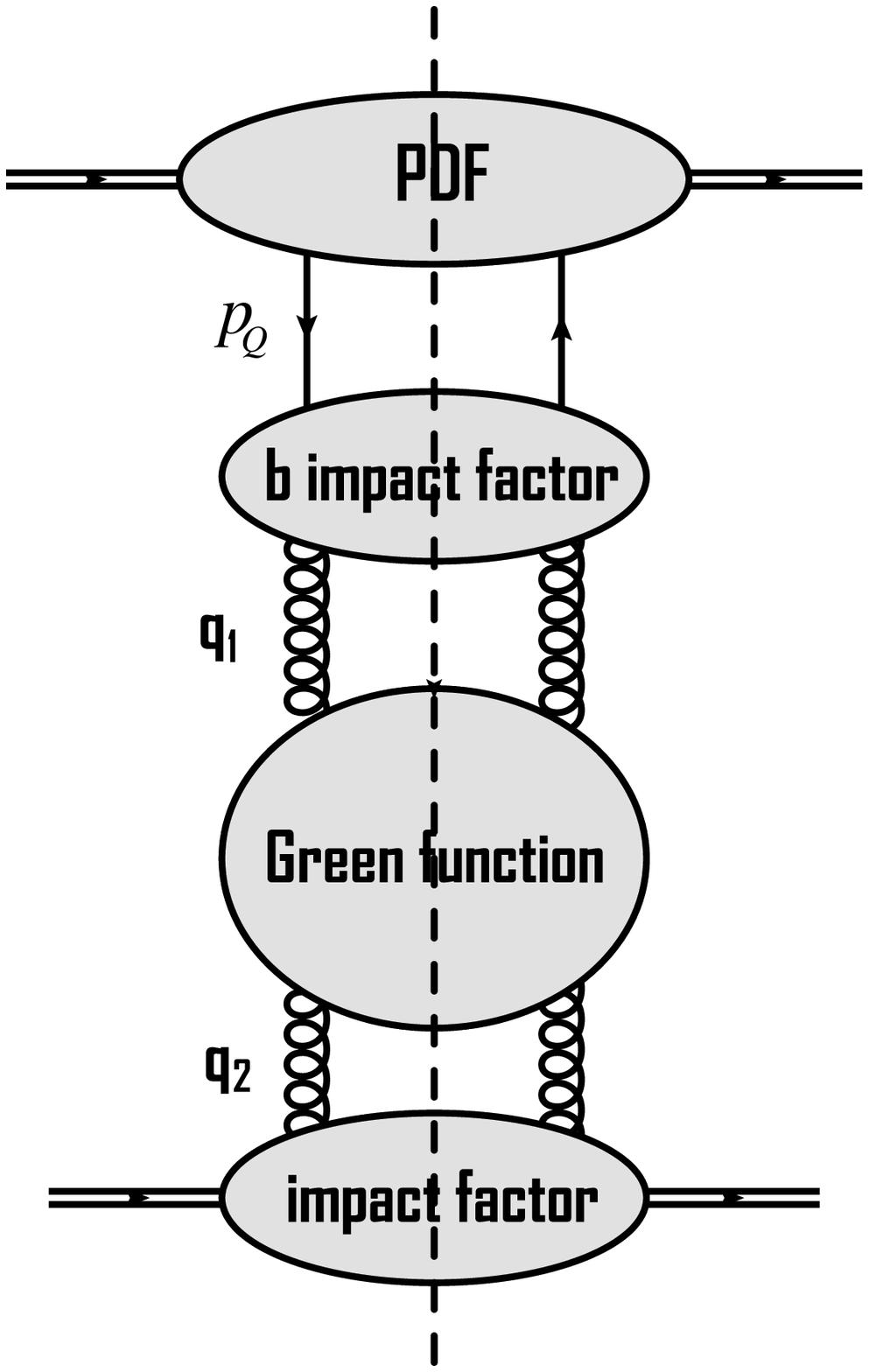}
    }    
     \end{picture}
  \begin{picture}(0,0)
    \put(222, -220){
      \includegraphics{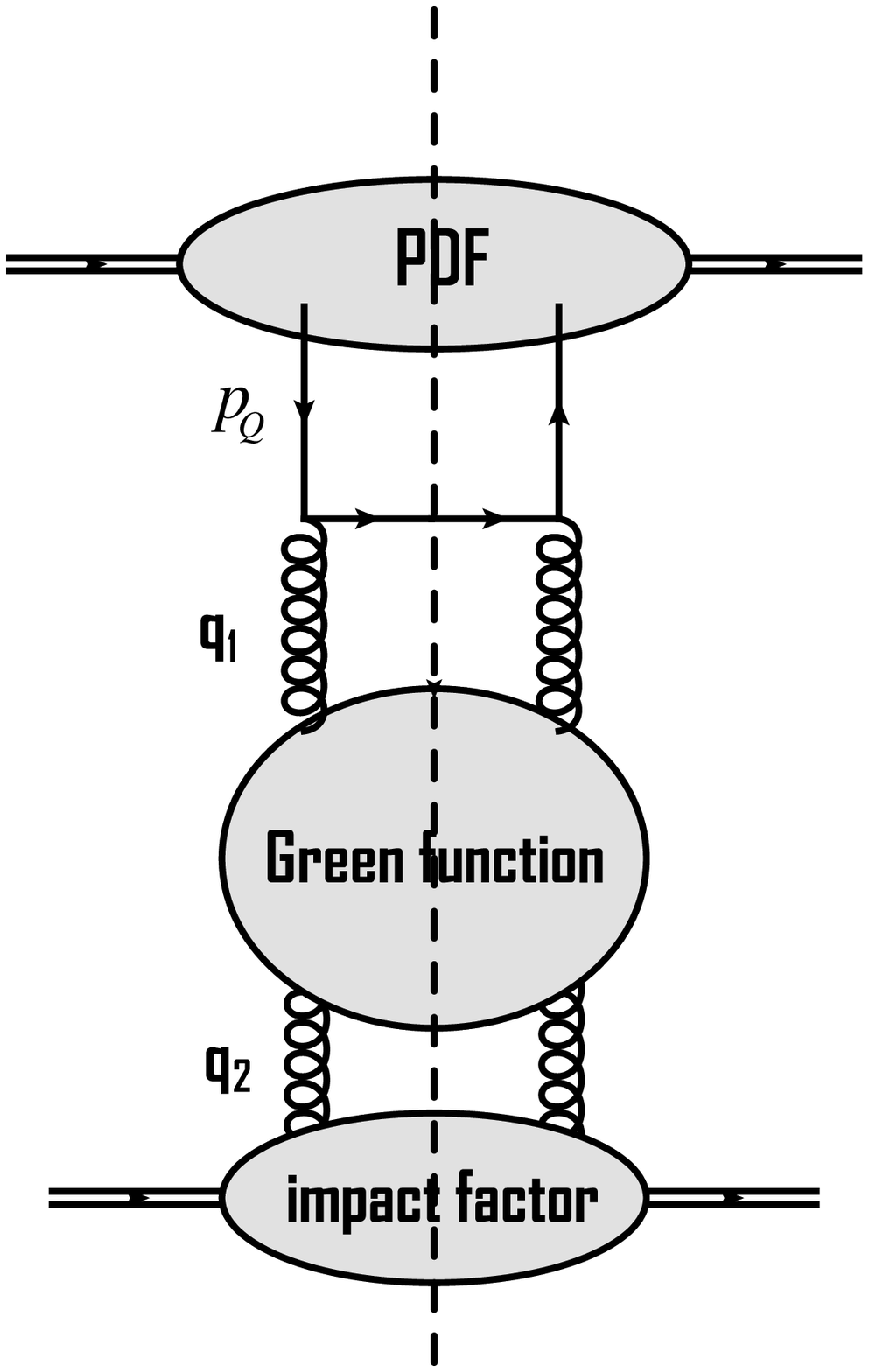}
    }    
     \end{picture}
\vspace{6cm}
\caption{Single bottom quark production in high energy
  factorization. The cross-section is given by a convolution of the
  gluon Green's function, the proton impact factor (at the bottom of
  the diagram) and the bottom quark impact factor.  A generic order
  heavy quark impact factor is depicted to the left whereas to the
  right the L$x$ impact factor is shown.}
\label{fig:diagrams}
\end{figure}

In the following we will study for typical LHC center-mass-energies $\sqrt{s}=8$ and~$13$ TeV the process 
\begin{align}
  \label{eq:1}
  \text{proton}(p_1) + \text{proton}(p_2) \to \text{bottom quark jet}(k) + X
\end{align}
where the jet rapidity is assumed to be close to the forward region
of the scattering proton with momentum $p_1$. We assume in the following light-like proton momenta $p_1$ and $p_2$, with $2 p_1 \cdot p_2 = s$. For the above process, the bottom quark jet
provides  a hard scale, both through the bottom mass $m_b$
and its transverse momentum $k_T$, which allows for an an analysis of this
process within QCD perturbation theory. 
 Furthermore, since the
scattering proton with momentum $p_2$ is separated from both the heavy
quark jet and the proton with momentum $p_1$ by a large interval in
rapidity, a description of the process within high energy
factorization is possible.
For sufficiently high $k_T$, this process is then described  at leading order through the partonic process $Q + g \to Q'$ convoluted with corresponding  gluon and heavy quark distribution functions. Within high energy factorization, the initial heavy quark is always taken at large $x_Q \sim 1$, while the gluon is pushed into the small $x_g \ll 1$ region, with the opposite configuration ($x_Q \ll 1$ and $x_g \sim 1$) suppressed by powers of the center-of-mass energy.
For the further analysis within high energy factorization (which includes a resummation a large terms $({\alpha}_s \ln 1/x_g)^n \sim 1$ to all orders in $\alpha_s$), it is then sufficient to analyze the process
\begin{align}
  \label{eq:3}
  Q(x_Q\cdot p_1) + p(p_2) \to \text{bottom quark jet}(k) + X',
\end{align}
{\it i.e.} we study scattering of a heavy quark on a proton together with production of a heavy quark jet in high energy limit,  see Fig.~\ref{fig:diagrams}. The cross-section for
for this process $\sigma_{Q}$ can be written as a
convolution of three objects:  the partonic heavy quark impact factor, the gluon
Green's function, which is a process independent universal quantity
and the proton impact factor.
Formally, this means that we can write   for the forward bottom quark cross-section:
\begin{equation}\label{eq:prot-HQ}
\sigma_{Q}(x_g, Q^2) = \frac{1}{(4 \pi)^4} \int \frac{d^2 \bm{q}_1}{{\bm q}_1^2} 
\int \frac{d^2 \bm{q}_2}{{\bm q}_2^2} \Phi_Q({ q}_1, Q^2) \mathcal{F}^{\text{DIS}}(x,  q_1, q_2)
\Phi_{p}( q_2, Q_0^2), 
\end{equation}
where  $Q$  is the hard scale related to the final state heavy quark momentum. We have introduced $q_{i}=\sqrt{{\bm q}_i^2}$, $i=1,2$ (the transverse momenta of $t$-channel gluons, see Fig.~\ref{fig:diagrams}). 
In the equation~\eqref{eq:prot-HQ} $\Phi_Q(q_1, Q^2) $ is the heavy quark impact factor, $\Phi_p(q_2,
Q_0^2)$ the proton impact factor and $ \mathcal{F}^{\text{DIS}}(x,
q_1, q_2)$ the gluon Green's function adapted for DIS-like kinematics.
$\Phi_Q(q_1, Q^2) $ and $ \mathcal{F}^{\text{DIS}}(x, q_1, q_2)$ are
quantities, that are calculable in perturbative QCD whereas
$\Phi_p(q_2, Q_0^2)$ is an object of  intrinsic non-perturbative
nature and has to be modeled.  We will use in this study the fit of 
Refs.~\cite{Hentschinski:2012kr,Hentschinski:2013id} which achieves a 
successful description of $F_2$ and $F_L$ HERA data with a very simple ansatz for the proton impact factor with three independent parameters.

When the two scales $Q^2$ and $Q_0^2$ are similar in size,
 the gluon Green's function
${\cal F}$ -- which is obtained as  the solution to the BFKL equation --
can be written at leading order as
\begin{eqnarray}
{\cal F}^{\text{Lx}}\left(s, q_1 , q_2\right) &=& \frac{1}{2 \pi q_1 \, q_2} \int \frac{d \omega}{2 \pi i} \int \frac{d \gamma}{2 \pi i} 
\left(\frac{q_1^2}{q_2^2}\right)^{\gamma-\frac{1}{2}} \left(\frac{s}{q_1 \, q_2}\right)^\omega \frac{1}{\omega- \bar{\alpha}_s \chi_0\left(\gamma\right)},
\label{GGF1}
\end{eqnarray}
with $\bar{\alpha}_s = \alpha_s N_c / \pi$ and $\chi_0(\gamma) = 2
\psi(1) - \psi(\gamma)-\psi(1-\gamma)$ the eigenvalue of the L$x$ BFKL
kernel with $\psi(\gamma)$ is the logarithmic derivative of the Euler
Gamma function. The gluon Green's function is universal and resums
$\bar{\alpha}_s^n \log^n{s}$ terms to all-orders in the strong
coupling.
 
In our setup however, $Q^2 \gg Q_0^2$ and this expression 
should be written in a form consistent with 
the resummation of $\bar{\alpha}_s \log{(1/x)}$ contributions:
\begin{eqnarray}
{\cal F} (s, q_1 , q_2) &=& \frac{1}{2 \pi q_1^2} \int \frac{d \omega}{2 \pi i} \int \frac{d \gamma}{2 \pi i} 
 \left(\frac{{q_1^2}}{q_2^2}\right)^{\gamma} \left(\frac{s}{q_1^2}\right)^\omega \frac{1}{\omega- \bar{\alpha}_s \chi_0\left(\gamma - \frac{\omega}{2}\right)}.
\end{eqnarray}
In the limits $\gamma \to 0,1$, the zeros of the denominator of the
integrand generate all-orders terms not compatible with DGLAP
evolution~\cite{Salam:1998tj,Vera:2005jt}. By taking into account the
NL$x$ correction to the BFKL kernel, the first of these pieces (${\cal
  O} (\alpha_s^2)$) is removed. Higher orders though however remain
and are numerically important. A scheme to eliminate these spurious
contributions was introduced in~\cite{Salam:1998tj} by using a
modified BFKL kernel in Eq.~(\ref{GGF1}) incorporating the change
$\chi_0(\gamma) \to 2 \psi(1) - \psi(\gamma + \tfrac{\omega}{2}) -
\psi(1-\gamma + \tfrac{\omega}{2}) $.

The NL$x$ kernel after collinear improvements can very well be
approximated  by breaking the transcendentality of the NL$x$ kernel and
solving it pole by pole and summing up the different solutions.  This
procedure was introduced in Ref.~\cite{Vera:2005jt} and we refer the
reader there for further details. The NL$x$ kernel with collinear
improvements we will be using hereafter reads
 \begin{equation}\label{eq:gluongf}
\chi\left(\gamma\right)={\bar\alpha}_s\chi_0\left(\gamma\right)+
{\bar\alpha}_s^2\chi_1\left(\gamma\right)-\frac{1}{2}{\bar\alpha}_s^2
\chi_0^{\prime}\left(\gamma\right)\chi_0\left(\gamma\right)+
\chi_{RG}({\bar\alpha}_s,\gamma,a,b).
\end{equation}
with
\begin{eqnarray}
\label{eq:RG}
&& \chi_{RG}(\bar{\alpha}_s, \gamma, a, b)
\,\, =  \,\,\bar{\alpha}_s (1+ a \bar{\alpha}_s) \left(\psi(\gamma) 
- \psi (\gamma-b \bar{\alpha}_s)\right) \nonumber\\
&& \qquad \quad - \frac{\bar{\alpha}_s^2}{2} \psi'' (1-\gamma)  - b \bar{\alpha}_s^2 \frac{\pi^2}{\sin^2{(\pi \gamma)}}
+ \frac{1}{2} \sum_{m=0}^\infty \Bigg(\gamma-1-m+b \bar{\alpha}_s  \nonumber\\
&&
\qquad \quad - \frac{2 \bar{\alpha}_s (1+a \bar{\alpha}_s)}{1-\gamma+m}
+ \sqrt{(\gamma-1-m+b \bar{\alpha}_s)^2+ 4 \bar{\alpha}_s (1+a \bar{\alpha}_s)} \Bigg).
\end{eqnarray}
For the NL$x$ BFKL kernel we have:
\begin{eqnarray}
\label{eq:chi1NLO}
\chi_1 (\gamma) & =& {\cal S} \chi_0 (\gamma) - \frac{\beta_0}{8 N_c}   \chi_0^2(\gamma)
+ \frac{ \Psi ''(\gamma) + \Psi''(1-\gamma)- \phi(\gamma)-\phi (1-\gamma) }{4} \nonumber  \\
&& \hspace{-1.8cm} - \frac{\pi^2 \cos{(\pi \gamma)}}{4 \sin^2{(\pi \gamma)}(1-2\gamma)}
\left[3+\left(1+\frac{n_f}{N_c^3}\right) \frac{2+3\gamma(1-\gamma)}{(3-2\gamma)(1+2\gamma)}\right] 
+ \frac{3}{2} \zeta(3) ,
\end{eqnarray}
with ${\cal S} =\tfrac{1}{3}-\tfrac{\pi^2}{12} + \tfrac{5 \beta_0}{12 N_c}$,  $\beta_0 =  \left(\frac{11}{3}N_c - \frac{2 }{3} n_f \right)$ and
\begin{eqnarray}
\phi(\gamma) + \phi (1-\gamma) &=&\sum_{m=0}^\infty \left(\frac{1}{\gamma+m}+\frac{1}{1-\gamma+m}\right)
\left[\Psi'\left(1+\frac{m}{2}\right)-\Psi'\left(\frac{1+m}{2}\right)\right],
\end{eqnarray}
whereas  the coefficients $a$ and $b$ read
\begin{eqnarray}
a &=& \frac{5}{12} \frac{\beta_0}{N_c} - \frac{13}{36} \frac{n_f}{N_c^3}- \frac{55}{36}, \hspace{1cm} 
b ~=~ - \frac{1}{8} \frac{\beta_0}{N_c} - \frac{n_f}{6N_c^3}- \frac{11}{12}.
\end{eqnarray}
To achieve a model with  sensible parameters for the
proton impact factor dominated by the infrared region, the Brodsky-Lepage-Mackenzie
(BLM) optimal scale setting scheme~\cite{Brodsky:1982gc}
has been used in
~\cite{Hentschinski:2012kr,Hentschinski:2013id} to fix the renormalization scale.\footnote{The first application of the BLM scheme was in
Ref.~\cite{Brodsky:1998kn,Brodsky:1997sd,Brodsky:1996sg} in the context of virtual photon-photon
scattering.} The BLM procedure is a way of absorbing the 
non conformal terms of the perturbative series in a redefinition 
of the coupling constant, to improve the convergence of the perturbative series.
Practically, one needs to extract the 
$\beta_0$-dependent part of an observable and 
choose the renormalization scale such that this part
vanishes. In the current case this leads to
\begin{eqnarray}
\label{eq:BLM}
\tilde{\alpha}_s \left(Q Q_0, \gamma \right) &=& \frac{4 N_c}{\beta_0 \left[\log{\left(\frac{Q Q_0}{ \Lambda^2}\right)} 
+\frac{1}{2} \chi_0 (\gamma) - \frac{5}{3} +2 \left(1+ \frac{2}{3} Y\right)\right]},
\end{eqnarray}
where we are using the momentum space (MOM) physical renormalization scheme based on a symmetric triple gluon vertex~\cite{Celmaster:1979km} with 
$Y \simeq 2.343907$ and gauge parameter $\xi =3$ The 
modifications we need in the
BFKL kernel in order to introduce this new scheme are $\bar{\alpha}_s  \to \tilde{\alpha}_s \left(Q Q_0, \gamma\right)$ and $\chi_1 (\gamma) \to \tilde{\chi}_1 (\gamma)$ in Eqs.~\eqref{eq:gluongf},~\eqref{eq:chi1NLO} together with the corresponding adjustments for the coefficients $a, b \to \tilde{a}, \tilde{b}$ which enter Eq.~(\ref{eq:RG}). The modified quantities read
\begin{eqnarray}
\tilde{\chi}_1 (\gamma) &=& \tilde{\cal S} \chi_0 (\gamma) + \frac{3}{2} \zeta(3)
+ \frac{ \Psi ''(\gamma) + \Psi''(1-\gamma)- \phi(\gamma)-\phi (1-\gamma) }{4} \nonumber \\
&-& \frac{\pi^2 \cos{(\pi \gamma)}}{4 \sin^2{(\pi \gamma)}(1-2\gamma)}
\left[3+\left(1+\frac{n_f}{N_c^3}\right) \frac{2+3\gamma(1-\gamma)}{(3-2\gamma)(1+2\gamma)}\right] \nonumber\\
&+&\frac{1}{8} \left[\frac{3}{2} (Y-1) \xi
   +\left(1-\frac{Y}{3}\right) \xi ^2+\frac{17 Y}{2}-\frac{\xi ^3}{6}\right] \chi_0 (\gamma), \\
\tilde{a} &=&  - \frac{13}{36} \frac{n_f}{N_c^3}- \frac{55}{36} + \frac{3 Y - 3}{16}\xi + \frac{3 - Y}{24} \xi^2 - \frac{1}{48}\xi^3 + \frac{17}{16}Y
 \\
\tilde{b} &=& - \frac{n_f}{6N_c^3}- \frac{11}{12},
\end{eqnarray}
where $\tilde{\cal S} =\tfrac{ (4-\pi^2)}{12}$. 
In addition, in order to access the region of small photon virtualities, in \cite{Hentschinski:2012kr,Hentschinski:2013id}, a parametrization of the running coupling introduced by Webber in 
Ref.~\cite{Webber:1998um} has been used, 
 \begin{eqnarray}
\alpha_s \left(\mu^2\right) =  \frac{4\pi}{\beta_0\ln{\frac{\mu^2}{\Lambda^2}}}
+ f\left(\frac{\mu^2}{\Lambda^2}\right) , \;\;\;\; f\left(\frac{\mu^2}{\Lambda^2}\right) =  \frac{4\pi}{\beta_0}\; \frac{ 125\left(1 + 4 \frac{\mu^2}{\Lambda^2}\right)}{\left(1 - \frac{\mu^2}{\Lambda^2}\right)\left(4 + \frac{\mu^2}{\Lambda^2}\right)^4},
\end{eqnarray}
with $\Lambda=0.21\;$GeV. At  low scales this modified running coupling  is consistent with global data of power corrections to perturbative observables,  while  for larger values it coincides  with the conventional perturbative running coupling constant.

Let us add here, that in a future analysis we plan to investigate
effects related to the choice of the renormalization scale and the
choice of the parametrization of the running of the strong coupling
(see Ref.~\cite{Chachamis:2013bwa} and also
Refs.~\cite{Rodrigo:2000pk,Rodrigo:1993hc,Rodrigo:1997zd}).

\section{The differential cross-section with bottom mass effects included}\label{sec:diff-cs}

As already mentioned in the previous section,
 the non-perturbative proton impact factor has to be modeled. We use here
 the same functional form as in Refs.~\cite{Hentschinski:2012kr,Hentschinski:2013id}:
\begin{eqnarray}
\label{eq:protoN}
\Phi_p \left(q,Q_0^2\right) &=& \frac{ {\cal C}}{2\pi\Gamma\left(\delta\right)} \left(\frac{q^2}{Q_0^2}\right)^\delta e^{-\frac{q^2}{Q_0^2}},
\end{eqnarray}
which introduces three independent free parameters and has a maximum at $q^2 = \delta \, Q_0^2$. Its  representation in $\gamma$ space reads
\begin{align}
\label{eq:imP}
  h_p(\gamma)&= \int \frac{d^2 {\bm q}}{\pi}  \Phi_p \left(q,Q_0^2\right) (q^2)^{-\gamma-1} =  \, {\cal C} \, \frac{\Gamma(\delta-\gamma)}{2\pi \Gamma\left(\delta\right)} (Q_0^2)^{-\gamma}.
\end{align}
The values of the parameters $Q_0, \delta$ and
$\mathcal{C}$ were determined from a fit to combined HERA
data. When the leading order photon impact factor was used the obtained values were 
$Q_0 = 0.28$ GeV, $\delta =8.4$ and $\mathcal{C} = 1.50$ whereas
in the case of the  kinematically improved photon impact
factor the last two change to $\mathcal{C} = 2.35$ and $\delta = 6.5$ with the number of flavors fixed to  $n_f=4$. We use both
sets of values for  $\mathcal{C}$ and $\delta$ in our numerical study later. Combining the BFKL Green's function for DIS kinematics  and the proton impact factor, we obtain the following expression for an  unintegrated gluon density within our setup
\begin{align}
  \label{eq:Gugd_propose} 
  G(x, q_1  ) & = \int \frac{d     {\bm q}_2^2}{ {\bm q}_2^2} 
\mathcal{F}^{\text{DIS}} \left(x, q_1, q_2 \right) 
\Phi_p \left( q_2, Q_0^2 \right)\;.
\end{align}
To obtain the complete NL$x$ BFKL Green's function we need to add to Eq.~\eqref{GGF1} apart from the  NL$x$ correction to the BFKL eigenvalue, non-exponentiating NLx $\beta_0$ terms. Following the treatment of  Ref.\cite{Hentschinski:2013id} one obtains
\begin{equation}\label{eq:Gudg}
\begin{split}  
  G\left(x, q_1, Q\right) & = 
 \frac{1}{q_1^2}\int _{-\infty}^\infty  \frac{d \nu}{2 \pi^2}   \; \;    \frac{\mathcal{C}\cdot \Gamma(\delta - i \nu - \frac{1}{2})} {\Gamma(\delta)}  \; \cdot \;  \left(\frac{1}{x}\right)^{{\chi} \left(\frac{1}{2} + i\nu \right)}  \left( \frac{{q}_1^2}{Q_0^2} \right)^{\frac{1}{2} + i \nu}\,\times
 \\ 
&  
\Bigg\{1    + \frac{\bar{\alpha}_s^2\beta_0  \chi_0 \left(\frac{1}{2} + i \nu\right) }{8 N_c} \log{\left(\frac{1}{x}\right)} 
\Bigg[- \psi \left(\delta-\frac{1}{2}-i \nu\right)
-   \log \frac{{q}_1^2}{Q^2}\Bigg]\Bigg\}\;.
\end{split}
\end{equation}
In the DIS analysis $Q^2$ has been identified with the virtuality of
the photon. In the present study we use instead the transverse momentum of
the bottom quark, 
$Q = { k}_T$, 
with $k_T=\sqrt{{\bm k}^2}$ the modulus of the transverse momentum of the heavy quark.  The other obvious choice for $Q=\sqrt{{ k}_T^2 + m_b^2}$ causes only small differences in the results.

\begin{figure}[tbh]
\vspace{0cm}
  \begin{picture}(0,0)
    \put(40, -220){
      \includegraphics{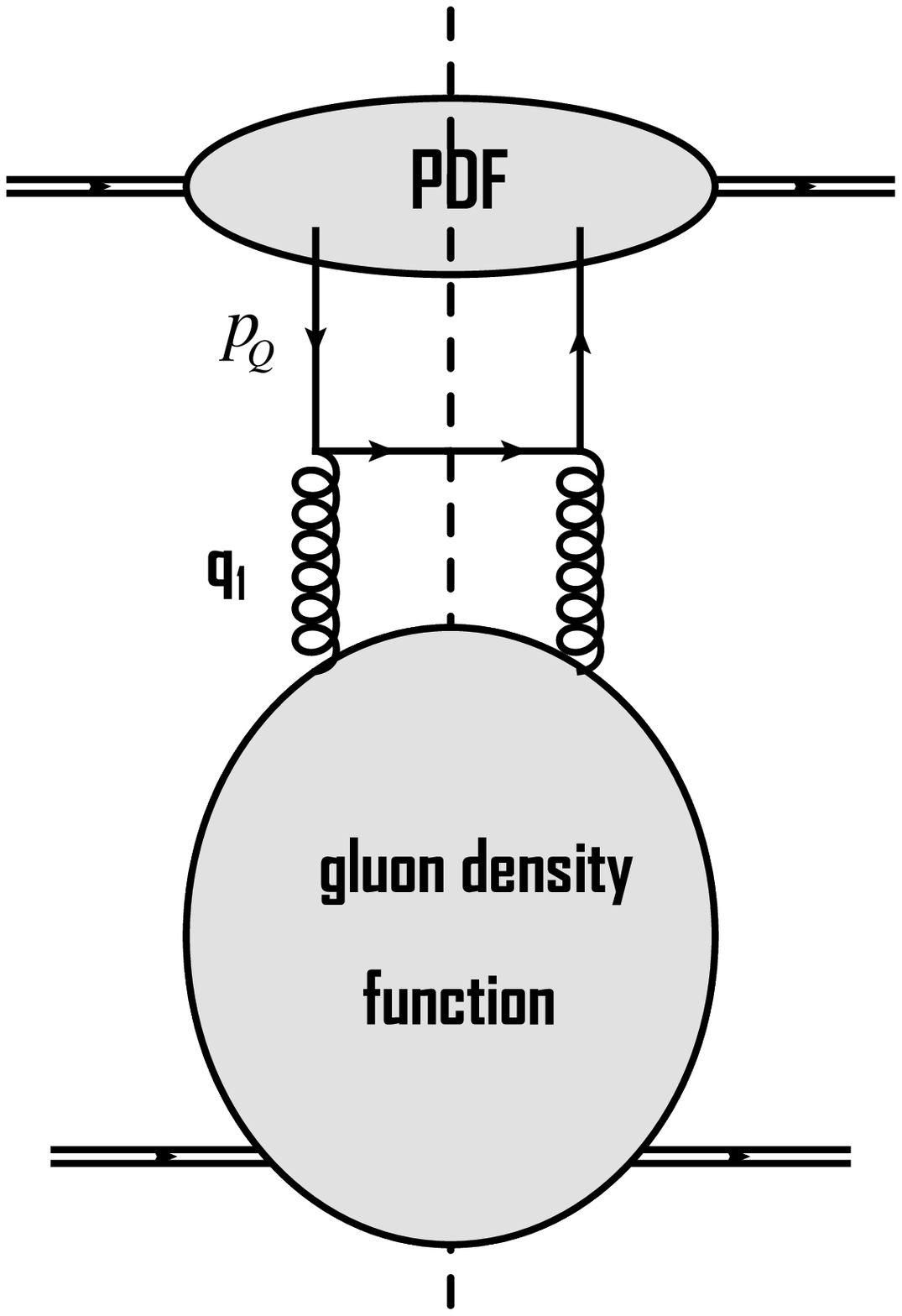}
    }    
     \end{picture}
  \begin{picture}(0,0)
    \put(222, -220){
      \includegraphics{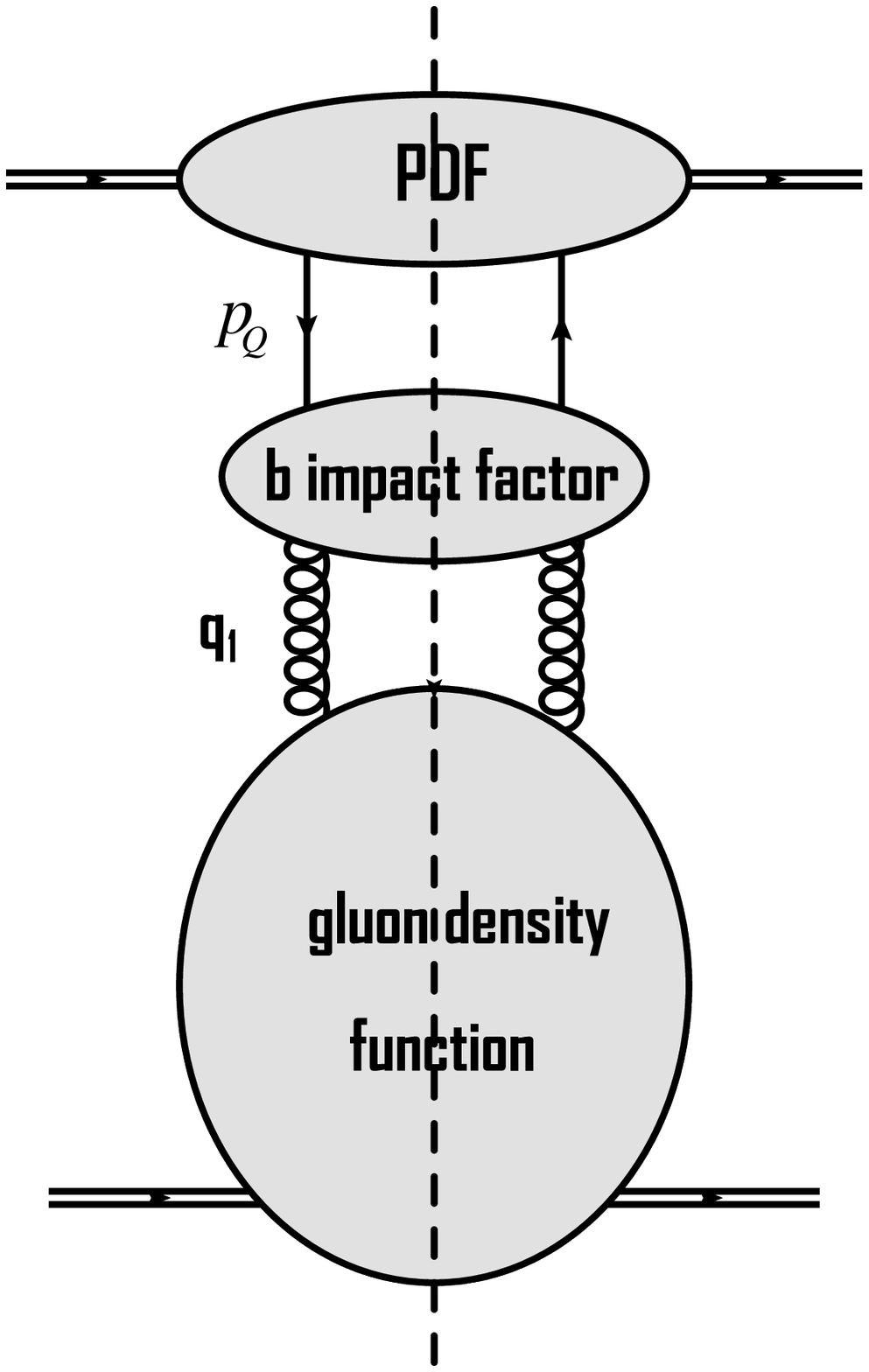}
    }    
     \end{picture}
\vspace{6cm}
\caption{The same process as in Fig.~\ref{fig:diagrams} here presented as a convolution
of the bottom quark impact factor and the gluon density.}
\label{fig:diagrams2}
\end{figure}

Once we have the formal definition of the gluon density given by
Eqs.~\eqref{eq:Gugd_propose} and~\eqref{eq:Gudg}, we can have an
alternative view at $Q p$-scattering depicted in
Fig.~\ref{fig:diagrams}.  In particular, we may consider it as a
convolution of the bottom quark impact factor with the gluon density
as shown in Fig.~\ref{fig:diagrams2}.
To obtain the complete forward heavy quark cross-section, we further require the bottom quark jet vertex  which at leading order
coincides with the massless quark impact factor modulo Dirac Delta functions to ensure momentum conservation. 
The heavy quark jet vertex at L$x$ depends therefore only implicitly
on the quark mass through the final state phase-space integration. We
should stress here, that a proper NL$x$ study with mass effects
properly introduced would be the desirable goal. Currently only the inclusive heavy quark impact factor, which describes the process $Q + g^* \to X''$, is available at NL$x$ accuracy \cite{Ciafaloni:2000sq,Chachamis:2013bwa}. While at leading order inclusive and jet impact factor coincide, a NL$x$ description of the process $Q + g^* \to \text{bottom quark jet} + X'''$ will depend explicitly on the details of the employed jet algorithm and hence differs from the corresponding inclusive result. 
In the current  study we therefore restrict ourselves  to the L$x$ jet vertex.  The
leading order impact factor is then obtained from the squared amplitude of the
subprocess $Q + g^* \to Q'$, integrated over the one-particle
invariant phase space $d \Phi^{(1)}$.  
The momenta of the incoming quark and gluon can be expressed in
Sudakov variables as $p_{Q} = x_Q p_1$ and $ p_g = x_g p_2 + {\bm q}_1
$ respectively.   Energy-momentum conservation identifies then
the gluon transverse momentum with the transverse momentum of the
final state bottom quark ${\bm q}_1 = {\bm k}$; the one-particle phase
space of the heavy quarks reads $d \Phi^{(1)} = 2 \pi \delta(x_Q x_g s
- {\bm k}^2 - m_b^2)$.  After a bit of Algebra, we obtain
for the $Q + g^* \to Q'$ partonic cross-section:
$$\hat{\sigma} =\sigma_0  \delta(x_Q x_g s - {\bm k}^2 - m_b^2) \;, \qquad \sigma_0 = \frac{\alpha_s 2 \pi^2}{N_c}  \;.$$
The total $p p \rightarrow Q + X$ cross-section then reads
 \begin{align}
   \label{eq:sigma-tot-1}
   \sigma_{p p \rightarrow Q + X} & = 
   \int_0^1 dx_Q  \int_0^1 \frac{dx_g}{x_g} \int \frac{d^2 {\bm k}}{\pi} \hat{\sigma}\cdot   \left[   f_Q(x_Q, \mu_f) +  f_{\bar{Q}}(x_Q, \mu_f) \right] G(x_g, {\bm k}_T, Q)\;.
 \end{align}
with $f_{i}$, $i=Q,\bar{Q}$ the collinear (anti-) bottom quark distribution and $\mu_f$ the collinear factorization scale. Fixing  $x_g =( {\bm k}^2 + m_b^2)/x_Q s$ and 
introducing the rapidity of the produced bottom quark $\eta = \frac{1}{2} \ln \frac{x_Q}{x_g}$,  the total
cross-section is recast into
\begin{equation}\label{eq:sigma-tot-2}
\begin{split}
  \sigma_{p p \rightarrow Q + X} 
 & =    \int\limits_{-\infty}^\infty  d\eta \int\frac{d^2 {\bm k}}{\pi}    \frac{ \sigma_0}{x_Q x_g s}   \,x_Q\left[ f_Q(x_Q, \mu_f)+ f_{\bar{Q}}(x_Q, \mu_f) \right])\, G(x_g, { k}_T, Q)\\
 & =    \int\limits_{-\infty}^\infty d\eta  \int \frac{d^2 {\bm k}}{\pi}    \frac{ \sigma_0}{{k}_T^2 + m_b^2} \,  x_Q \left[ f_Q(x_Q, \mu_f)+ f_{\bar{Q}}(x_Q, \mu_f)\right]\, G(x_g, {k}_T, Q)\;.
\end{split}
\end{equation}
 After integrating  over the
azimuthal angle of ${\bm k}$, the $p p \rightarrow Q' + X$ double differential cross-section finally  reads
\begin{align}
  \label{eq:sigma-diff-1}
 \frac{d \sigma_{p p \rightarrow Q + X}}{ d\eta~ dk_T}  & =       \frac{2 k_T\cdot \sigma_0}{k_T^2 + m_b^2}   \,x_Q\,\left[ f_Q(x_Q, \mu_f) + f_{\bar{Q}}(x_Q, \mu_f)\right]\, G(x_g, k_T^2, Q)\;.
\end{align}
with
\begin{equation}
\label{eq:2}
x_Q=e^{\eta}\sqrt{\frac{m_b^2+{\bm k}^2}{s}}\;,\qquad
x_g=e^{-\eta}\sqrt{\frac{m_b^2+{\bm k}^2}{s}}\;.
\end{equation}
Leaving aside for the time being the  dependence on the collinear bottom quark  distribution function $\,x_Q\,f_{i}(x_Q)$, $i=Q, {\bar{Q}}$ we will have
for the $Q p \to Q'$ cross-section in $\nu$-space
\begin{equation}\label{eq:crosssection2}
\begin{split}
\frac{d\sigma_{Qp \to Q'}}{d \eta d k_T}&=
 \;\frac{ \sigma_0 \cdot \mathcal{C}}{k_T\cdot (k_T^2+m_b^2)} \;  \int\limits_{-\infty}^{\infty}
\frac{d\nu}{ \pi^2}\;x^{-\chi\left(\frac{1}{2}+i\nu\right)}\;
\;
\frac{\Gamma\left(\delta-\frac{1}{2}-i\nu\right)}{\Gamma(\delta)}\;
 \left(\frac{k_T^2}{Q_0^2}\right)^{1/2+i\nu} 
\\\;
&\times \bigg\{
1
+
{\bar\alpha}_s^2 \log\left(\frac{1}{x}\right)\frac{\beta_0}{8N_c}\chi_0
\left(\frac{1}{2}+i\nu\right)\left[-
\psi ^{(0)}\left(\delta -i \nu -\frac{1}{2}\right) 
 \right]
\bigg\}\;.
\end{split}
\end{equation} 
 Fixing the factorization scale of the bottom quark PDF to $\mu_f = \sqrt{k_T^2 + m_b^2}$,  the double differential cross-section for single bottom quark production in proton-proton collisions reads
\begin{equation}\label{eq:xsectioneta}
\frac{d\sigma_{p p \rightarrow Q' + X}}{d\eta\,d k_T}=
\frac{d\sigma_{Qp \to Q'}}{d \eta\,d k_T}\;x_Q\bigg[f_Q\!\left(x_Q, \sqrt{k_T^2 + m_b^2} \right)+f_{\bar Q}\!\left(x_Q, \sqrt{k_T^2 + m_b^2} \right)\bigg]\;,
\end{equation}
which we use to produce all of our numerical results in the next section.

\section{Numerical results}\label{sec:numres}

In this section we present our predictions for the differential cross-section, $\eta$- and $k_T$-distribution, for single bottom quark/anti-quark production
at the LHC. The analysis does not make distinction between the bottom quark and anti-quark.

\begin{figure}[tbh]
\vspace{-3.2cm}
  \begin{picture}(0,0)
    \put(13, -210){
      \includegraphics{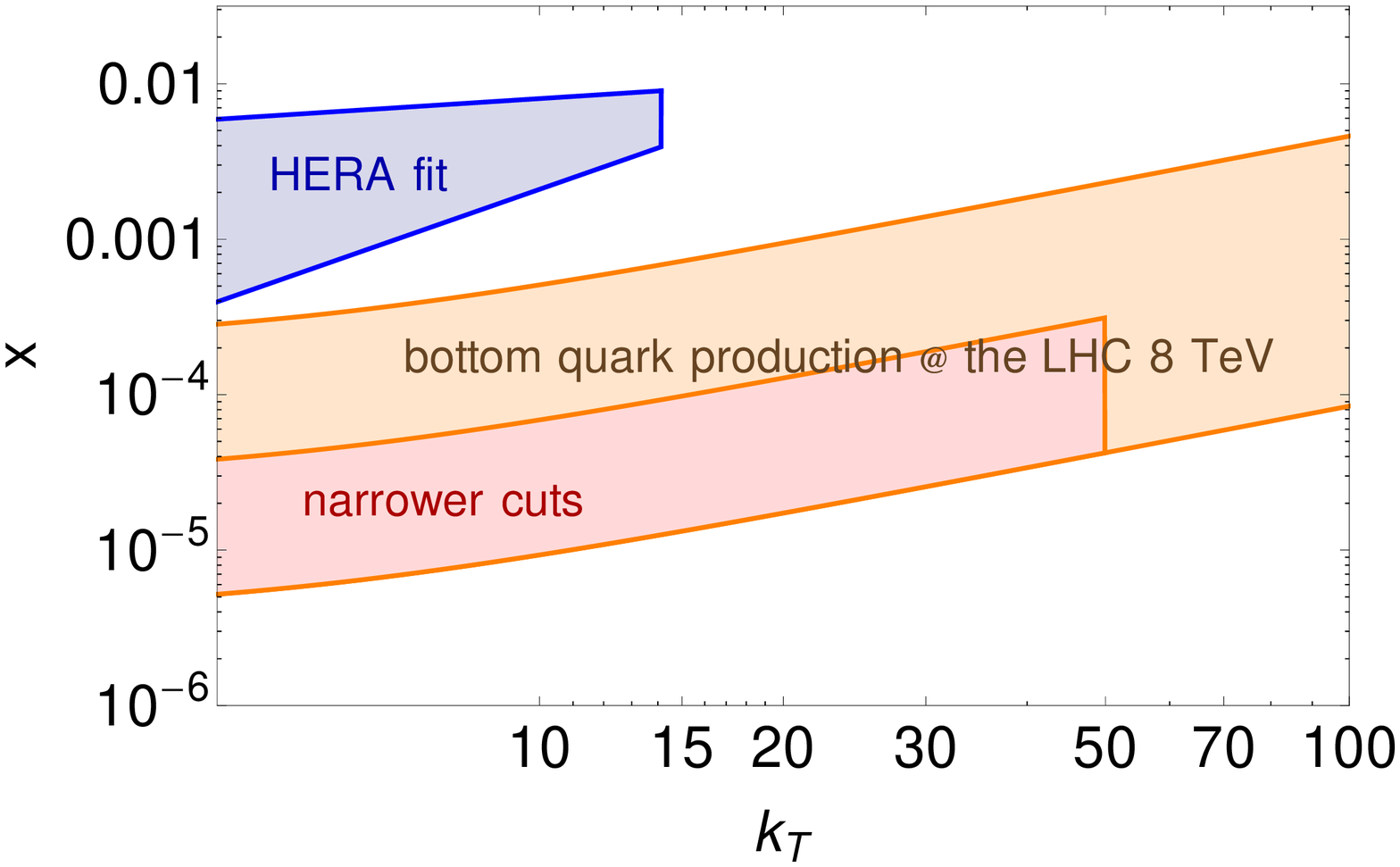}
    }    
     \end{picture}
  \begin{picture}(0,0)
    \put(215, -210){
      \includegraphics{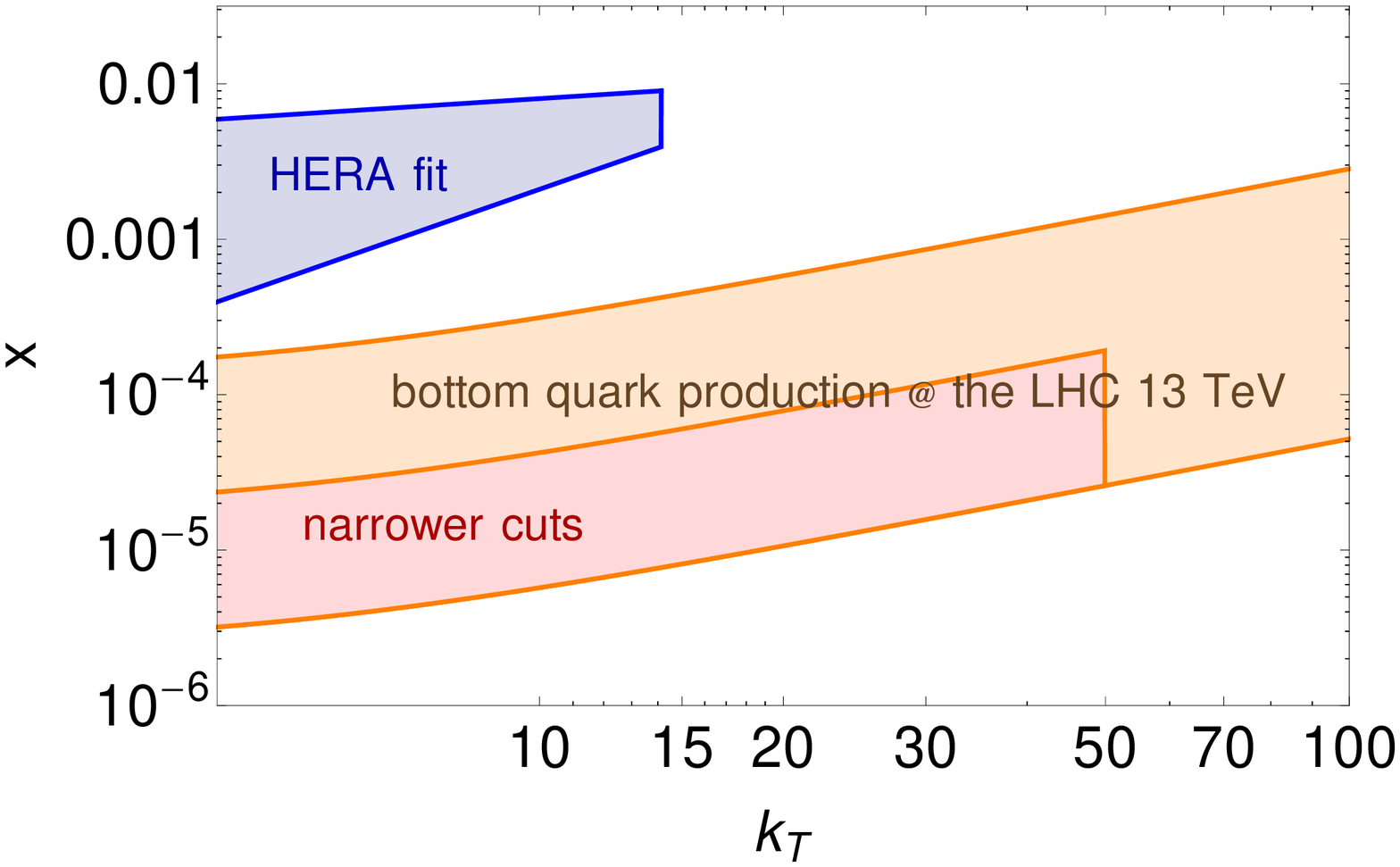}
    }    
     \end{picture}
\vspace{7.15cm}
\caption{Light blue area: kinematic region corresponding to the original fit of the proton impact factor.
Light orange area: phase space for the kinematic cuts  ($1<\eta<5$, $4\;$GeV$<k_T<100\;$GeV).
Light red area: phase space for the kinematic cuts ($3<\eta<5$, $4\;$ GeV$<k_T<50\;$ GeV).}
\label{fig:phasespace}
\end{figure}

In Fig.~\ref{fig:phasespace}, we show the range of maximal/minimal $x_g$ for each value of $k_T$. The light blue area
corresponds to the Bjorken $x$ and $Q$ ranges of the data, that were used for
the $F_2$ and $F_L$ fit in
~\cite{Hentschinski:2012kr,Hentschinski:2013id}. The light orange area
corresponds to the $x_g$ and $k_T$ ranges in our calculation of the
bottom quark cross-section when the rapidity takes values between 1
and 5 and $k_T$ is constrained to $4\;$GeV$<k_T<100\;$GeV.  Lastly,
the light red area corresponds to the $x$ and $k_T$ ranges when the
rapidity takes values between 3 and 5 and $k_T$ is constrained to
$4\;$GeV$<k_T<50\;$GeV. This kinematical range is not covered by the general purpose detectors ATLAS and CMS, but is accessible by the LHCb detector~\cite{Alves:2008zz} designed for these kind of measurements.
 
It is more than evident, that the kinematic
region we are covering for bottom quark production does not overlap at
all with the kinematic region in which the unintegrated gluon density
was obtained. In particular, the values of the unintegrated gluon
density tested in our setup are not directly constrained by HERA
data. They are rather calculated through evolving the results of the
HERA fit towards both smaller $x$ values and larger $k_T$ using our
collinear improved solution to the BFKL equation.  Nevertheless, as we
shall see, the unintegrated gluon density based on that specific
model, gives results very close to {\sc Pythia} 8.1. It has to be
stressed though, that the whole approach carries uncertainties which
at present cannot be quantified and one has to be cautious not to
interpret these high energy factorization results as the final word
within the BFKL approach.

In all the next figures, predictions with $\delta=8.4$ and ${\mathcal
  C}=1.5$ are plotted with solid red lines, the results with
$\delta=6.5$ and ${\mathcal C}=2.35$ are plotted with dashed red lines
and the results by {\sc Pythia} 8.1 are plotted with purple solid
lines.
We have used the MSTW 2008 NLO parton density functions~\cite{mstw2008} 
throughout the entire section. The bottom quark mass was set to $m_b=4.7\;$GeV for the high energy factorization result whereas in {\sc Pythia} 8.1
the program default value was used.


\begin{figure}[tbh]
\vspace{0cm}
  \begin{picture}(0,0)
    \put(3, -120){
      \includegraphics{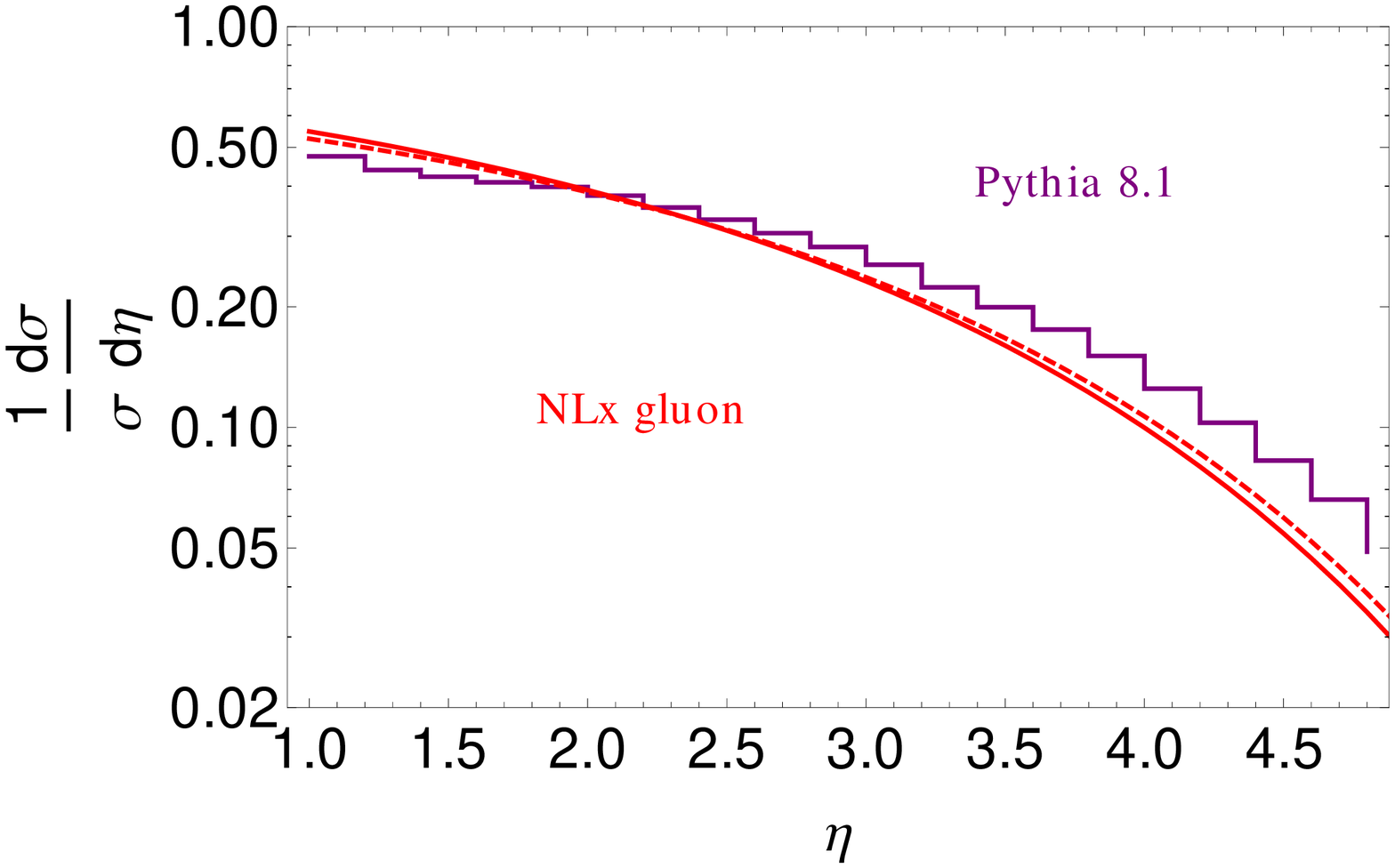}
    }    
     \end{picture}
  \begin{picture}(0,0)
    \put(205, -120){
      \includegraphics{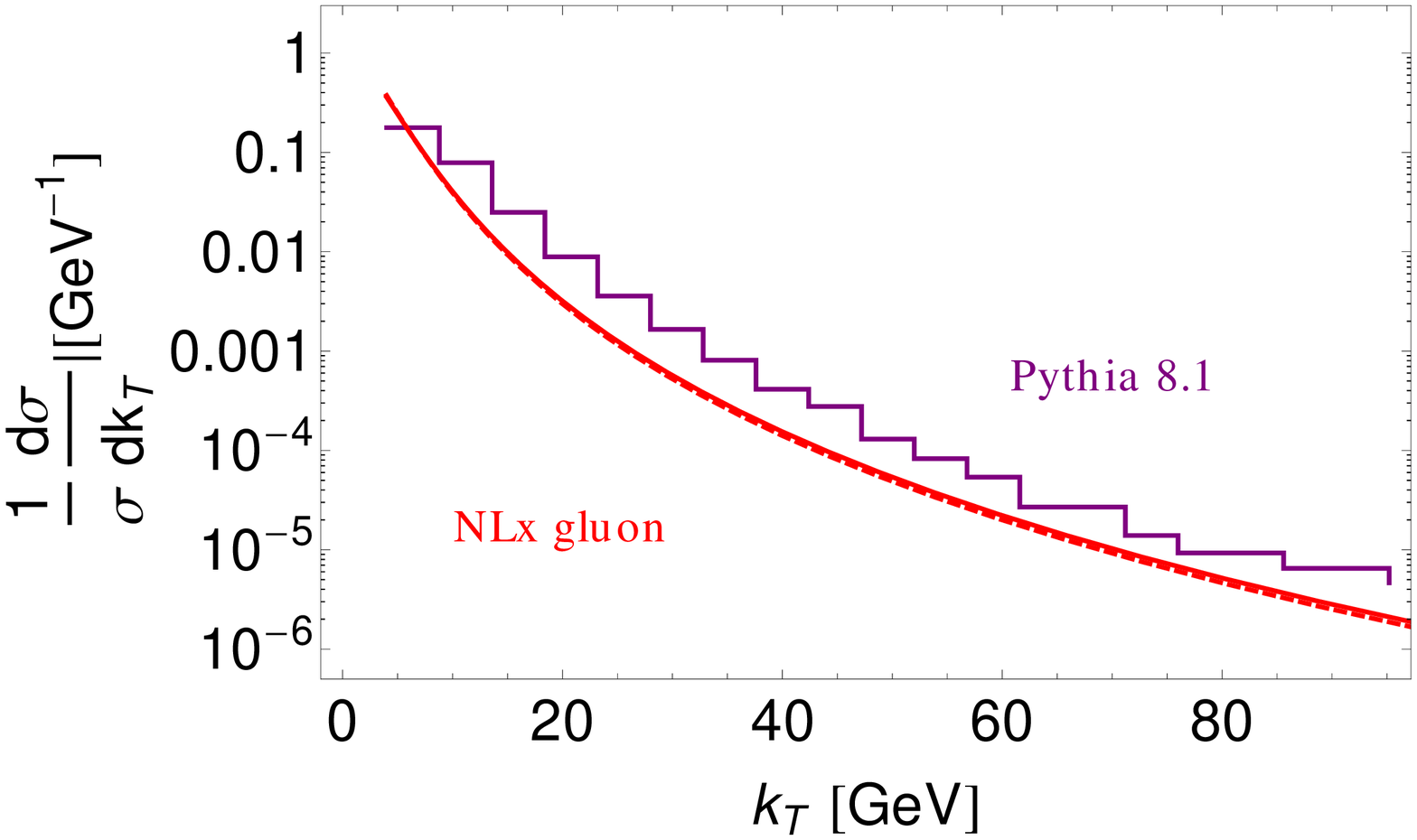}
    }    
     \end{picture}
\vspace{4.0cm}
\caption{Collision energy $\sqrt{s} = 8\;$TeV.
Left: $\eta$-distribution after integrating
over $k_T$ in the  range $4\;\textrm{GeV}<k_T<100\;\textrm{GeV}$. 
Right: $k_T$-distribution after integrating over $\eta$ in the range $1<\eta<5$.
Both distributions are  normalized by the 
integrated cross-section over $\eta$ and $k_T$ in the ranges $4\;\textrm{GeV}<k_T<100\;\textrm{GeV}$ and
$1<\eta<5$.}
\label{fig:results2-8}
\end{figure}

\begin{figure}[tbh]
\vspace{-3.4cm}
  \begin{picture}(0,0)
    \put(3, -210){
      \includegraphics{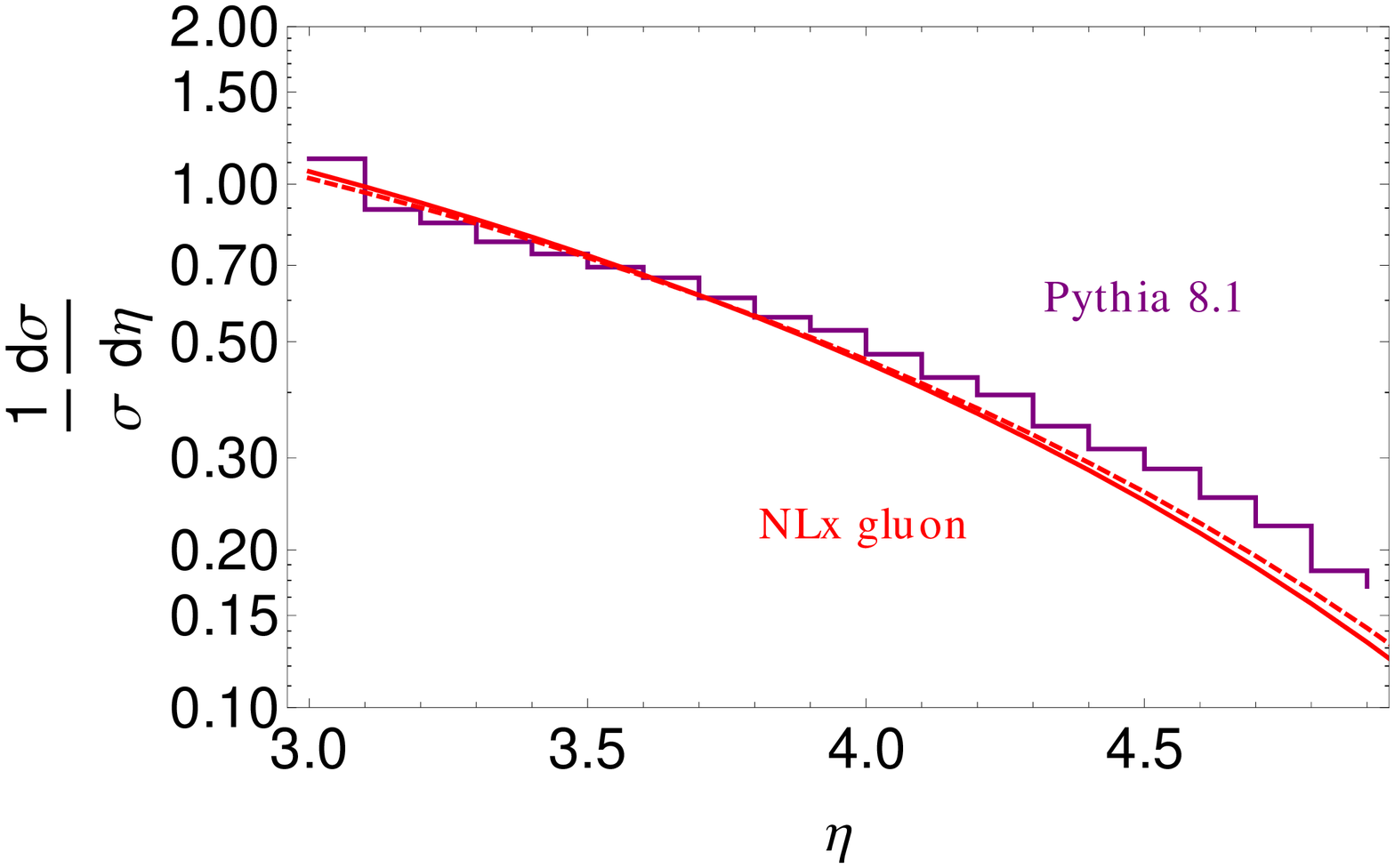}
    }    
     \end{picture}
  \begin{picture}(0,0)
    \put(205, -210){
      \includegraphics{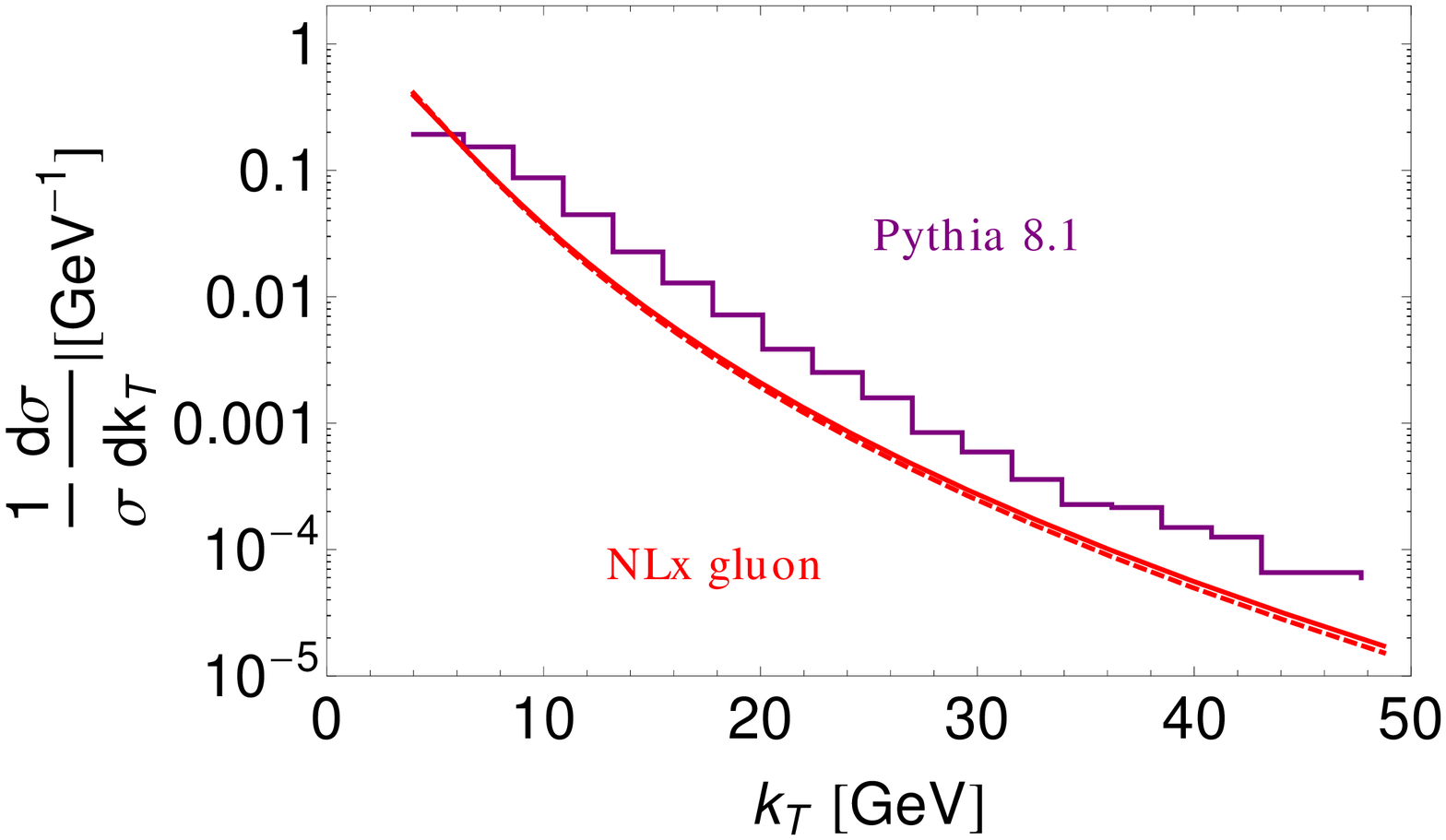}
    }    
     \end{picture}
\vspace{7.15cm}
\caption{Collision energy $\sqrt{s} = 8\;$TeV.
Left: $\eta$-distribution after integrating
over $k_T$ in the  range $4\;\textrm{GeV}<k_T<50\;\textrm{GeV}$. 
Right: $k_T$-distribution after integrating over $\eta$ in the range $3<\eta<5$.
Both distributions are  normalized by the 
integrated cross-section over $\eta$ and $k_T$ in the ranges $4\;\textrm{GeV}<k_T<50\;\textrm{GeV}$ and
$3<\eta<5$.}
\label{fig:results4-8}
\end{figure}
\begin{figure}[tbh]
\vspace{0cm}
  \begin{picture}(0,0)
    \put(3, -120){
      \includegraphics{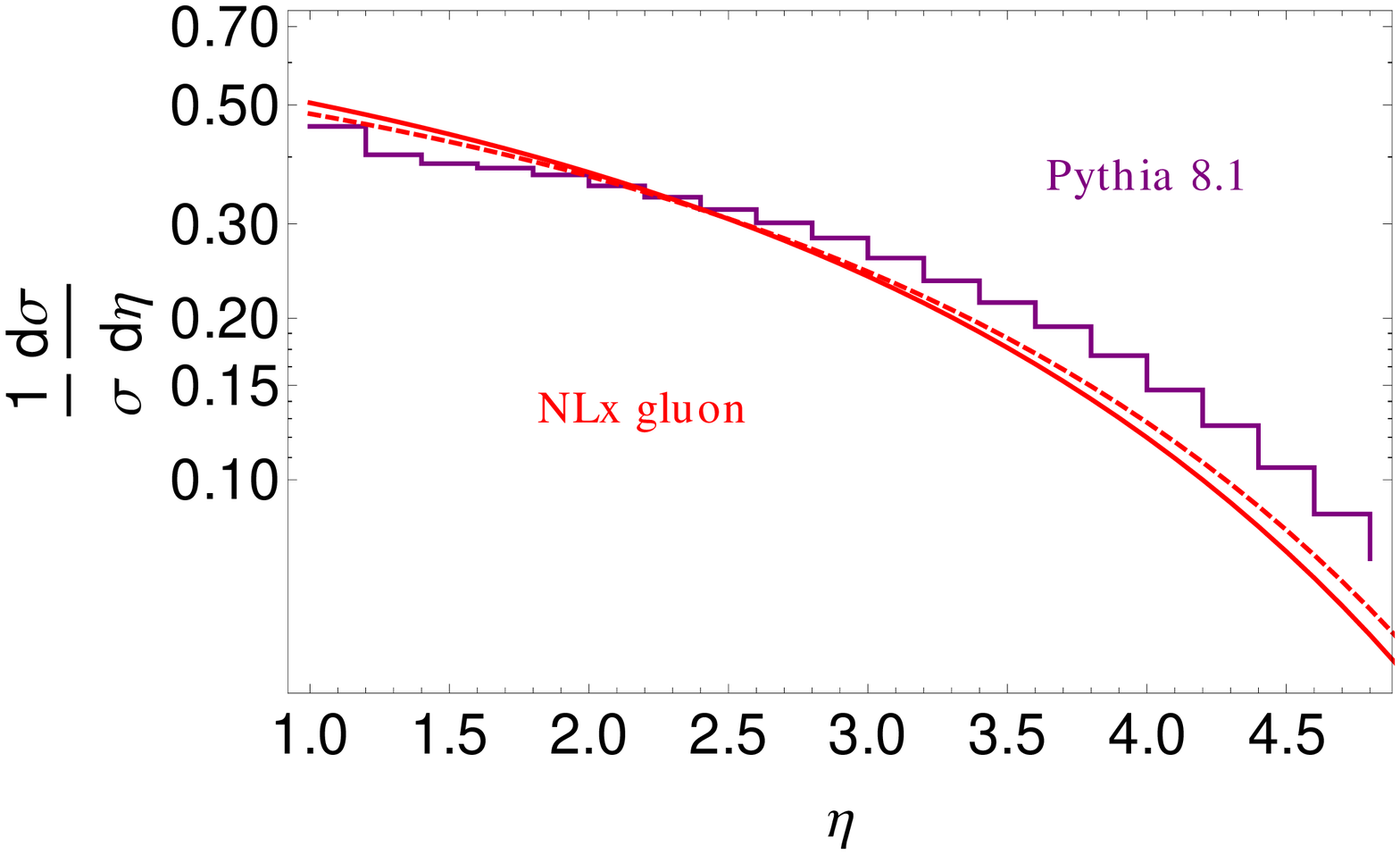}
    }    
     \end{picture}
  \begin{picture}(0,0)
    \put(205, -120){
      \includegraphics{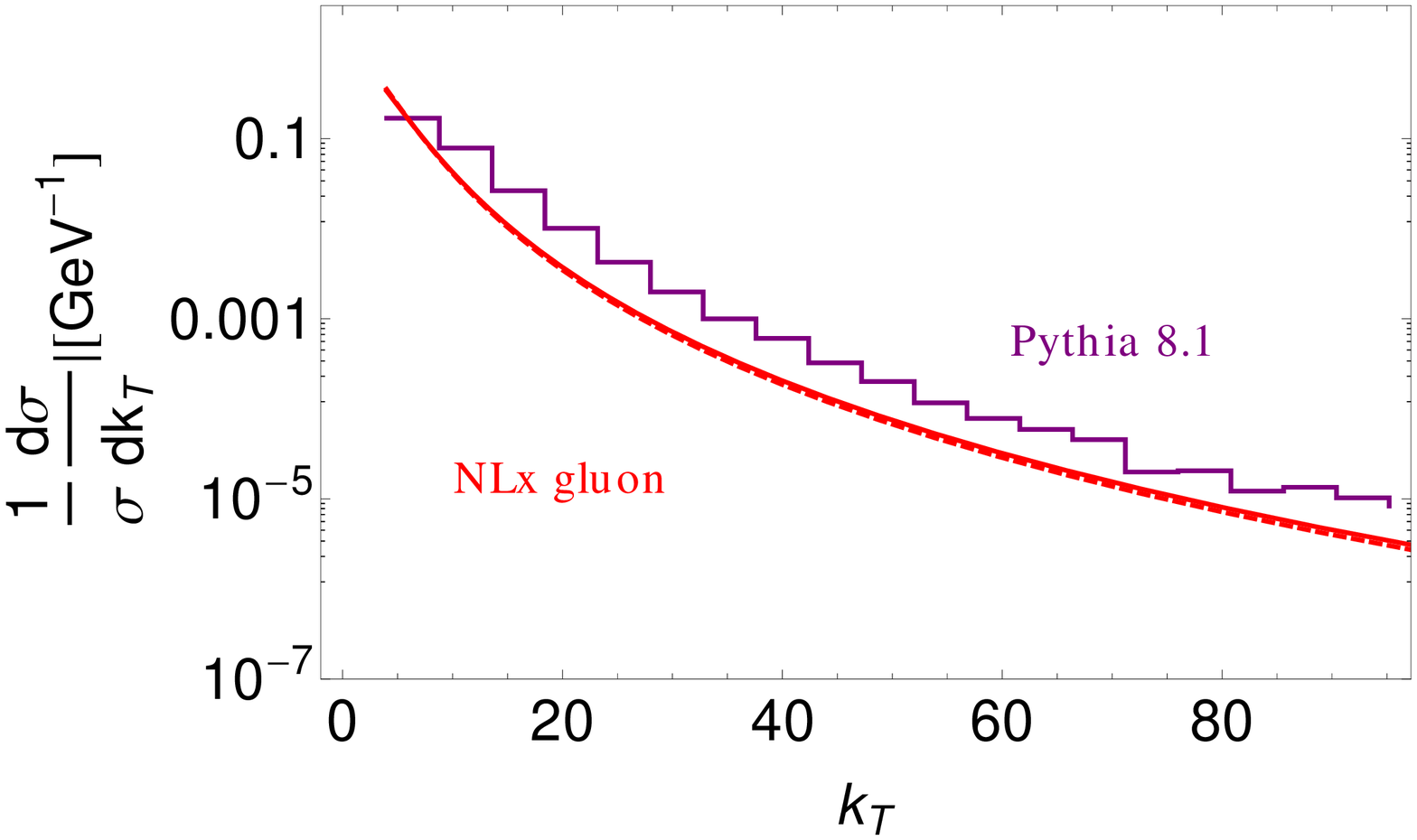}
    }    
     \end{picture}
\vspace{4.0cm}
\caption{Collision energy $\sqrt{s} = 13\;$TeV.
Left: $\eta$-distribution after integrating
over $k_T$ in the  range $4\;\textrm{GeV}<k_T<100\;\textrm{GeV}$. 
Right: $k_T$-distribution after integrating over $\eta$ in the range $1<\eta<5$.
Both distributions are normalized by the 
integrated cross-section
over $\eta$ and $k_T$ in the ranges $4\;\textrm{GeV}<k_T<100\;\textrm{GeV}$ and
$1<\eta<5$.}
\label{fig:results3}
\end{figure}
\begin{figure}[tbh]
\vspace{-3.4cm}
  \begin{picture}(0,0)
    \put(3, -210){
      \includegraphics{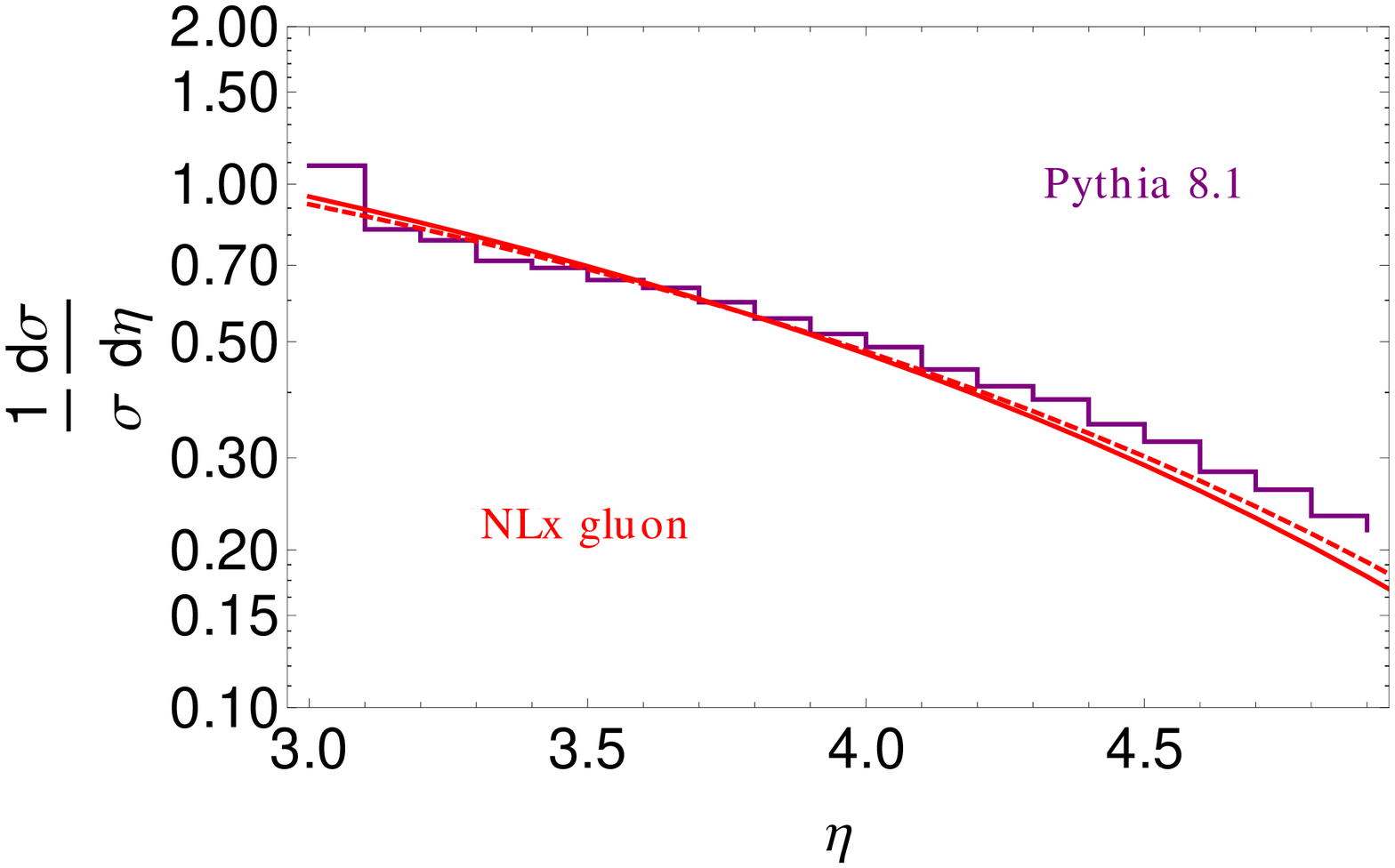}
    }    
     \end{picture}
  \begin{picture}(0,0)
    \put(205, -210){
      \includegraphics{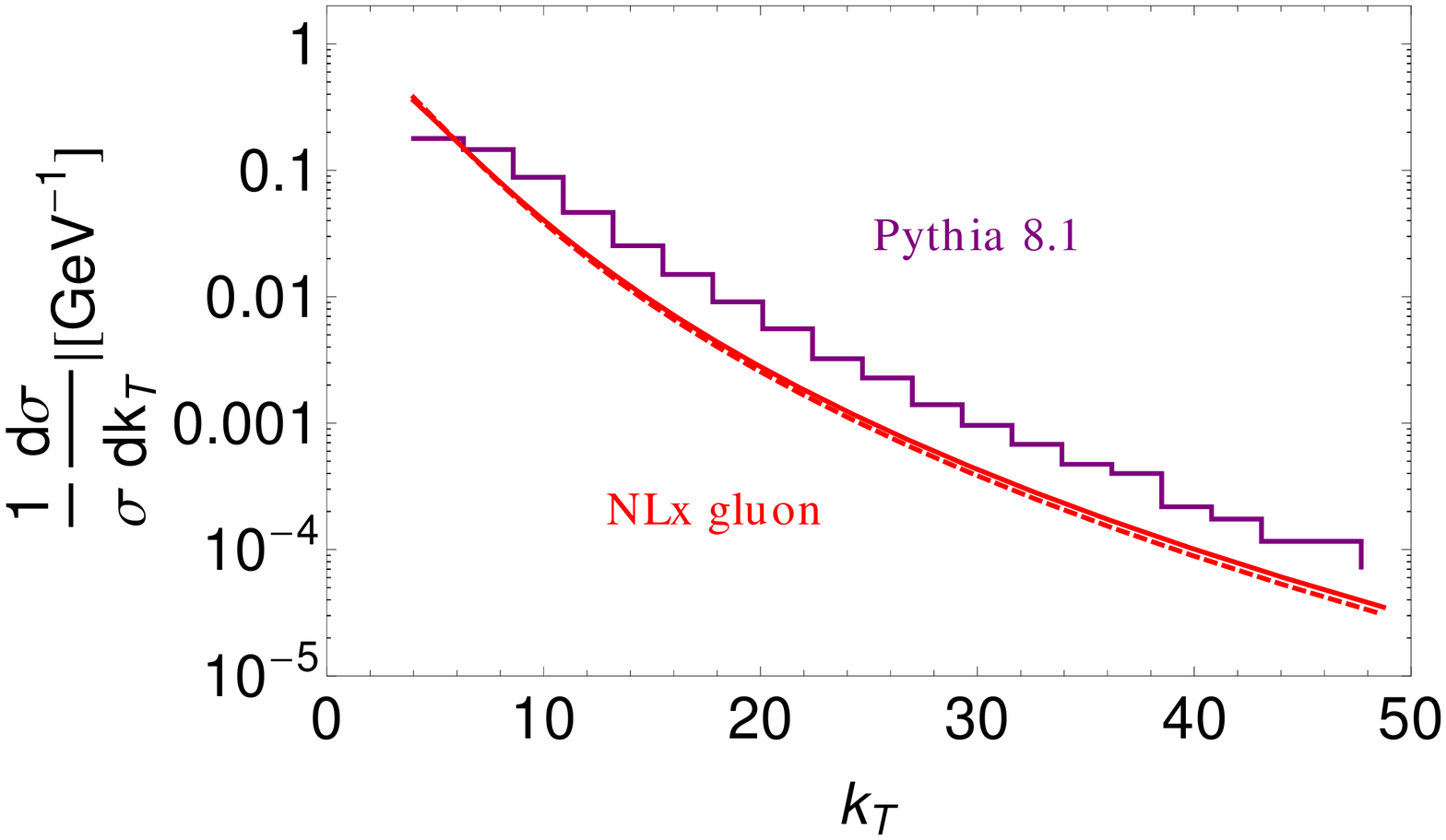}
    }    
     \end{picture}
\vspace{7.15cm}
\caption{Collision energy $\sqrt{s} = 13\;$TeV.
Left: $\eta$-distribution after integrating
over $k_T$ in the  range $4\;\textrm{GeV}<k_T<50\;\textrm{GeV}$. 
Right: $k_T$-distribution after integrating over $\eta$ in the range $3<\eta<5$.
Both distributions are  normalized by the 
integrated cross-section over $\eta$ and $k_T$ in the ranges $4\;\textrm{GeV}<k_T<50\;\textrm{GeV}$ and
$3<\eta<5$.}
\label{fig:results5}
\end{figure}
In Figs.~\ref{fig:results2-8} and ~\ref{fig:results4-8} we present results for $\sqrt{s} = 8$ TeV and in 
Figs.~\ref{fig:results3} and ~\ref{fig:results5} for $\sqrt{s} = 13$ TeV.
In Figs.~\ref{fig:results2-8} and~\ref{fig:results3} we integrate the differential cross-section~\eqref{eq:xsectioneta} over the rapidity of the quark in the range $1<\eta<5$ and over $k_T$ in the range $4$~GeV $<k_T<100$~GeV while in
Figs.~\ref{fig:results4-8} and~\ref{fig:results5} we integrate over the rapidity of the quark in the range $3<\eta<5$
and over the transverse momentum in the range $4$~GeV $<k_T<50$~GeV.
The differential distributions
are normalized by the integrated cross-section over the corresponding rapidity and $k_T$ ranges.

A first observation is that  when we compare
the $k_T$ distributions calculated in high energy factorization and by {\sc Pythia} 8.1 we see the former to be smaller than the latter due to the difference in the shape at small $k_T$. The shape of the $k_T$ distribution though is very similar at large $k_T$ in both approaches.

The main finding comes forward when we focus on the rapidity distributions.
The high energy factorization result seems to be somehow larger than the {\sc Pythia} 8.1 estimate
for small rapidities whereas for larger $\eta$
it drops faster than the   {\sc Pythia} 8.1 result and at the high end of the rapidity it lies below it.
The same trend is followed for both center-of-mass energies
and for both $k_T$ integration ranges ($4\;\textrm{GeV}<k_T<50\;\textrm{GeV}$ and
$4\;\textrm{GeV}<k_T<100\;\textrm{GeV}$).

Finally, let us note, that a change of values for the parameters 
$\delta$ and ${\mathcal C}$ from $\delta=8.4$ and 
${\mathcal C}=1.5$ to $\delta=6.5$ and ${\mathcal C}=2.35$ results to a
slightly smaller NL$x$ cross-section.

\section{Conclusions and Outlook}\label{sec:conclusions}

We have presented a study of the rapidity and $k_T$ (transversal momentum)
differential distributions for single bottom quark production
at the LHC calculated both in high energy factorization and by the Monte Carlo program {\sc Pythia} 8.1. 
and for $\sqrt{s} = 8$ and $13\;$TeV.
Within the former framework, we have used a model for the proton impact factor, the NL$x$
BFKL gluon Green's function and the L$x$ heavy quark jet vertex with bottom mass effects included.

The main result of our study concerns the rapidity distributions. The high energy factorization estimate
is for small rapidities larger than the {\sc Pythia} 8.1 result but as the rapidity
approaches some middle range value, the fall becomes steeper and the estimate gets smaller
than the {\sc Pythia} 8.1 estimate. The $k_T$-distributions have very similar shapes but the
{\sc Pythia} 8.1 result is almost always larger.

The calculation presented in this article suggests, within its limitations,
namely, that it is only a partial NL$x$ calculation and that the unintegrated gluon density is probed in a region not covered by the  $F_2$
and $F_L$ fit, that single bottom quark production might be 
used as an experimental probe  of the $k_T$-factorization scheme 
and the validity of  high energy factorization for LHC processes.
It also shows, that our initial assumption, that the unintegrated gluon
density from HERA would not fail at the LHC, was justified.
We plan to extend our study toward obtaining more
exclusive information with regard to the final state by using the BFKL
Monte Carlo code {\sc BFKLex}~\cite{bfklex,Chachamis:2011rw}. The code is an implementation
of an iterative solution to the NL$x$ BFKL equation and has already been
used in a number of
projects~\cite{Chachamis:2011nz,Chachamis:2012fk,Chachamis:2012qw}.
We expect, that once the NL$x$ massive quark jet vertex is ready, we will
able to present a more refined study of the process
and a detailed analysis on the BFKL predictions for the single bottom quark
cross section.

\section*{Acknowledgements}

M.~D. acknowledges support from Juan de la Cierva programme
(JCI-2011-11382). M.~H. acknowledges support by
UNAM-DGAPA-PAPIIT grant number 101515 and CONACyT-Mexico grant number
128534. G.~C. acknowledges support from Marie Curie Actions (PIEF-GA-2011-298582). A.S.V. acknowledges support from European Commission under contract LHCPhenoNet (PITN-GA-2010-264564), Madrid Regional Government (HEPHACOSESP-1473), Spanish Government (MICINN (FPA2010-17747)) and Spanish MINECO Centro de Excelencia Severo Ochoa Programme (SEV-2012-0249). This work has been supported in part by the Spanish Government
and ERDF funds from the EU Commission [Grants No. FPA2011-23778,
FPA2014-53631-C2-1-P No. CSD2007-00042 (Consolider Project CPAN)] and
by Generalitat Valenciana under Grant No. PROMETEOII/2013/007. This work has been supported in part by the Ministry of Economy and Competitiveness (MINECO), under grant number FPA2013-44773-P

\end{document}